\documentclass{PHYEAUTH} %used for Physica E
\usepackage{graphicx}
\usepackage{amsmath}
\usepackage{amssymb}

\newcommand{\bra}[1]{\left\langle#1\right|}
\newcommand{\ket}[1]{\left|#1\right\rangle}

\newcommand{\abs}[1]{\left|#1\right|}
\begin{document}

\begin{frontmatter}

\title{
Isospin Blockade in Transport through Vertical Double Quantum Dots
}

\author[address1]{Bernhard Wunsch\thanksref{thank1}},
\author[address2]{David Jacob}
and
\author[address1]{Daniela Pfannkuche}
\address[address1]{I. Institute of Theoretical Physics, University of
  Hamburg, Jungiusstr. 9, D--20355 Hamburg, Germany}
\address[address2]{Departamento de F\'isica Aplicada, Universidad de
  Alicante, San Vicente del Raspeig, 03690 Alicante, Spain}
\thanks[thank1]{
Corresponding author. 
E-mail: bwunsch@physnet.uni-hamburg.de}
\begin{abstract}
  We study the spectrum and the transport properties of two
  identical, vertically coupled quantum dots in a perpendicular
  magnetic field.  We find correlation-induced energy crossings in a
  magnetic field sweep between states differing only in the vertical
  degree of freedom. Considering the influence of a slight asymmetry
  between the dots caused by the applied source-drain voltage in vertical
  transport experiments these crossings convert to anticrossings
  accompanied by the build-up of charge polarization which is tunable
  by the perpendicular magnetic field. The polarization strongly
  affects the vertical transport through the double quantum dot and is
  manifest in an isospin blockade and the appearance of negative
  differential conductances in the magnetic field range where the
  charge localization occurs.
\end{abstract}
\end{frontmatter}
Double quantum dots have recently attracted much interest as possible
implementations of qubits and as model systems to study the
few-particle physics of molecular binding \cite{Isospin,Rontani,Pi}.
In vertically coupled double quantum dots (DQD) the molecular binding
for a single electron is determined by the vertical degree of freedom,
while the lateral degree of freedom reflects the physics of the single
dots or quasiatoms building up the molecule. In this paper we describe
level crossings between few-particle states differing only in parity
along the vertical direction.  These crossings are based on Coulomb
correlations between the electrons and may lead to strong charge
polarization, which is tunable by an external perpendicular magnetic
field\cite{Jacob}. Thus, these crossings mark changes of the molecular
binding which originate from many-body effects. Besides its importance
for quantum dot physics, the effect illuminates the analogy between
real spin and isospin, the latter of which describes the vertical
degree of freedom\cite{Isospin}. This analogy manifests itself in an
isospin blockade of transport through the double dot
system\cite{Weinmann,Tokura,Partoens}.
Furthermore, the localization provides an effective two-level
system which may be useful as a field-tunable charge qubit.

We model the DQD in vertical direction by two parallel layers
separated by a distance $d$. In lateral direction the electrons are
confined by a rotationally-symmetric parabolic potential of strength
$\hbar \omega_0$. Therefore, the vertical
degree of freedom is reduced to an additional spin-like degree of
freedom, the isospin\cite{Isospin}. In analogy to the real spin one can
define a spin algebra for the isospin, where the z-component $I_z$
specifies the vertical degree of freedom.  An electron with $I_z =
+1/2$ ($I_z = -1/2$) is situated in the upper (lower) dot.  We perform
an exact diagonalization of the few-particle Hamiltonian including tunneling
between the dots as well as the Zeeman term and the full Coulomb
interaction. Thus, correlations between electrons are fully
taken into account.  The strength of tunneling $t$ is specified by the
energy splitting between the symmetric and antisymmetric single particle
states in z-direction, $t=-\Delta_{SAS}/2$. Due to the symmetries of
the system the z-component of the angular momentum, $M$, the magnitude
of the total spin, $S$, and the z-component of the spin, $S_z$ are
conserved.
If both dots are identical and no asymmetry between the layers is
present, an additional symmetry leads to the conservation of the
isospin parity, $\hat{P} = 2^{N_e}\cdot \hat{I}_x^{(1)} \otimes \ldots
\otimes \hat{I}_x^{(N_e)}$ specified by the quantum number $P=\pm 1$.
The isospin parity flips all isospins, e.g.  it moves all electron
orbitals from the upper dot to the lower dot and vice versa.
Increasing the vertical magnetic field effectively leads to a stronger
lateral confinement of the electrons. Therefore, the Coulomb energy
increases with the magnetic field which causes ground state (GS) crossings
between states differing in angular momentum and/or spin~\cite{Isospin}.

In this paper we discuss energy crossings in a magnetic field sweep
between states differing only in parity, e.g. in the vertical degree
of freedom \cite{Jacob}.
These
crossings originate from the different scaling with magnetic field of the
\emph{intradot} Coulomb interaction that accounts for the interaction
between electrons on the same dot and \emph{interdot} Coulomb
interaction acting between electrons on different dots.  The intradot
interaction increases faster with increasing magnetic field than the
interdot interaction, that is limited to $1/d$~\cite{Isospin}. Hence,
the Coulomb correlations in the eigenstates of the DQD are magnetic
field dependent.

In the following we study parity crossings in a DQD containing three
electrons. For small tunneling $t\ll\hbar\omega_0$ the energy
splitting between the parity eigenstates $\ket{P=\pm 1}$ within the
same set of quantum numbers ($M,S,S_z$) is due to their different
tunneling energies, as the occupation probabilities of symmetric and
antisymmetric orbitals depends on parity. However due to the
magnetic-field-dependent Coulomb correlations the favored parity may
change with magnetic field.  In the example discussed here $P=-1$
is preferred for magnetic fields $B<7.75$~T whereas $P=1$ is
preferred for strong magnetic fields, $B>7.75$~T.

\begin{figure}
 \includegraphics[scale=0.8, angle=0]{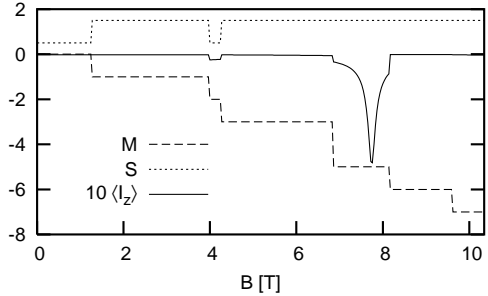}
  \caption{
    Angular momentum $M$, total spin $S$ and expectation value of the
    z-component of the isospin $\langle I_z \rangle$ for the
    three-electron GS. The peak in $\langle I_z \rangle$
    illustrates the charge localization that corresponds to a parity
    anticrossing (see text).  Parameters~\cite{material}: Tunneling
    strength: $t = -0.059~{\rm meV}$, confinement: $\hbar \, \omega_0
    = 2.96~{\rm meV}$, layer separation: $d = 19.6~{\rm nm}$, asymmetry: $V_z =
    5.9*10^{-4}$~meV.  }
  \label{fig:localization}
\end{figure}

Parity crossings between states with the same quantum numbers $M, S$
and $S_z$ can have drastic effects. Fig.~\ref{fig:localization}
illustrates the characteristics of the three-electron ground state
(GS) in a slightly asymmetric DQD as function of magnetic field. An
asymmetry between the dots is modeled by a small voltage drop between
them.  We account for it by adding the term $\hat{\mathcal{V}}_z = V_z
\cdot I_z$ to the Hamiltonian, where $I_z=\sum_{i=1}^{3} I_z^{(i)}$ is
the z-component of the total isospin and $V_z$ is assumed to be much
smaller than all other parameters $V_z \ll t, \hbar \omega_0$.  At
around $B=7.75$~T there is a sharp peak in the average isospin
z-component, corresponding to a strong charge polarization. Right at
the peak, two electrons occupy the lower dot and only one electron
occupies the upper one.  The charge localization is a consequence of a
parity crossing in the symmetric DQD, that turns into an anticrossing
due to the slight asymmetry between the dots. The width of the peak
and the energy splitting at the anticrossing are proportional to the
strength of asymmetry\cite{Jacob}. For the calculation presented in
Fig.~\ref{fig:localization} we chose the parameters such, that the
anticrossing occurs in the GS of a DQD containing three electrons.
But the effect of charge localization is neither
bound to this particular choice of parameters nor to three electron
states.  We found similar effects also for 5 electrons and/or in other
subsets of quantum numbers $M, S$ and $S_z$.

In the following we show that the charge localization drastically
affects the transport through the DQD. We assume a vertical transport
setup where the current and the differential conductance through the
DQD are measured as response to a source-drain voltage $V_{sd}$ and a
gate voltage $V_G$ \cite{Pi,Austing,Kouwenhoven}. The source-drain
voltage determines the width of the transport window where electrons
from the source reservoir can tunnel through the DQD to the drain
reservoir, whereas the gate voltage changes the energies of the DQD
proportionally to $V_G N$ where N denotes the number of electrons in
the DQD. The energy needed for a transition between an $N$-electron
state $i$ with energy $E(N,i)$ and an $N+1$-electron state $j$ with
energy $E(N+1,j)$ is therefore given by
$\mu(N+1,j;N,i)=E(N+1,j)-E(N,i) - e \alpha V_G$, where $\alpha$
denotes a constant proportionality factor.  Hence the gate voltage
together with the eigenenergies of the DQD determine which transitions
lie within the transport window.  On the other side the value of the
transition rates depends on the form of the eigenstates. In special
cases a transition is forbidden since it violates conservation laws. A
well-known example is the spin-blockade that occurs if the spins of
the two states differ by more than $\Delta S =\pm \frac{1}{2}$
\cite{Weinmann}. For general cases the effect of the structure of the
wavefunctions enters the transition rates through the spectral weights
\cite{Pfannkuche} that determine the probability for an electron
entering a $N$-electron state of the DQD to cause a transition to a
$N+1$-electron state of the DQD.

In our calculations we assume the coupling of the DQD to the external
reservoirs to be weak in comparison with the average energy spacing in
the DQD. Consequently the coupling to the external reservoirs is
treated to lowest order perturbation theory and the transport through
the DQD is described by sequential-tunneling processes in and out of
many-particle eigenstates of the isolated DQD \cite{Pfannkuche}.
Therefore, tunneling causes transitions between those eigenstates of
the DQD which differ by one in the number of electrons. We assume that
an electron in the upper (lower) reservoir can only tunnel into the
upper (lower) dot, which leads to different spectral weights for the
tunneling-in and tunneling-out process \cite{Pfannkuche}.  Assuming
the source reservoir to be the upper reservoir, the spectral weights
for a transition between the $i$-th two-electron state and the $j$-th
three-electron state caused by a tunneling-in (-out) process is given
by $\sum_{l} \abs{\bra{N_e=2,i} d_{l \pm} \ket{N_e=3,j} }^2$.  $l$
runs over all possible single particle states and $d_{l +}$ ($d_{l
  -}$) denotes an annihilator of an electron in orbital $l$ in the
upper (lower) dot. We include the asymmetry between the dots caused by
the source-drain voltage applied across the DQD by setting the
asymmetry $V_z=\frac{1}{20}e V_{sd}$.  With these assumptions we
calculate the stationary current and the differential conductance
through the DQD in a vertical transport experiment by solving the rate
equations containing the transition rates and the probabilities of
occupying the many particle eigenstate of the DQD \cite{Pfannkuche}.

\begin{figure}
 \includegraphics[scale=1, angle=0]{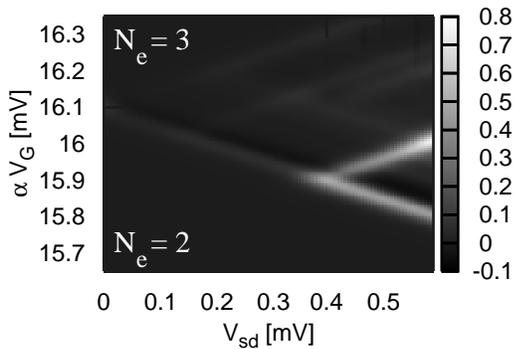}
  \caption{
    Section of charging diagram showing the differential conductance
    $G=\partial I / \partial U_{sd}$ for vertical transport through
    DQD.  $B=7.74$ T (i.e. three-electron GS is strongly polarized).
    The conductance is strongly suppressed at low transport voltages,
    due to an isospin blockade.  Units of $G$: $e \Gamma
    / mV$, where $\Gamma=DOS \abs{T_R}^2 \frac{2 \pi}{\hbar}$
    determines the coupling to the external reservoirs, DOS is the
    density of states in the reservoirs, $T_R$: tunneling matrix
    elements to the reservoirs. Parameters like in Fig.
    \ref{fig:localization} and temperature $T=140$~mK.  }
  \label{fig:CB_wc=2.245}
\end{figure}

Figure~\ref{fig:CB_wc=2.245} shows a charging diagram where the
differential conductance $G=\partial I / \partial V_{sd}$ is plotted
for each pair of transport voltage and gate voltage.  The conductance
shows significant values only for source-drain voltages $V_{sd} >
0.35$~mV. This is surprising since one could expect that the maxima of
the conductance form Coulomb diamonds starting from $V_{sd}=0$~mV and
$V_G=16.1$~mV, since at these voltages the Fermi energies of the
external reservoirs align with the energy needed for a GS-GS
transition between two and three electrons.  The results shown in
Fig.~\ref{fig:CB_wc=2.245} resemble the well-known spin
blockade\cite{Weinmann}. In fact our results can be explained by an
isospin blockade due to the charge polarization of the three electron
GS. An electron tunneling from the source reservoir (upper reservoir)
to the upper quantum dot can hardly cause a transition from the
unpolarized two-electron GS to the three-electron GS that has two
electron charges in the lower dot. In other words, the two electron
state is approximately an eigenstate to $I_z=0$, which by an electron
entering the upper dot turns approximately to an eigenstate to
$I_z=\frac{1}{2}$.  This state has no overlap with the three electron
GS, that is approximately an eigenstate to $I_z=-\frac{1}{2}$. This
explains why the spectral weight and therefore the transition rate for
the tunneling-in process nearly vanish and illuminates the relation to
the spin blockade, which is based on the orthogonality of eigenstates
with different spin.
Rising the source-drain voltage other transitions enter the transport
window that enable transport. The strong suppression of the current
discussed here demonstrates again the influence of the structure of
the eigenfunctions on the transport characteristics \cite{Pfannkuche}.

\begin{figure}
\includegraphics[scale=1, angle=0]{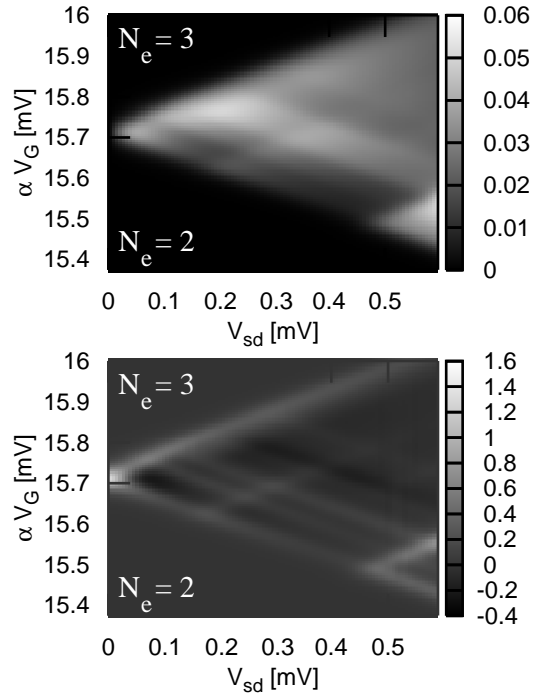}
\caption{
  Charging diagrams for transport through DQD. Asymmetry between the
  dots increases with increasing transport voltage reducing the
  current (upper part) and leading to broad regions of negative
  differential conductance (lower part) indicated by the dark black
  regions. Current in units of $e\Gamma$, $B = 7.24$~T; Other parameters like in Fig. \ref{fig:CB_wc=2.245}.}
\label{fig:CB_wc=2.1}
\end{figure}

Another prominent effect of charge localization on the transport
properties is the appearance of broad regions of negative differential
conductances in the charging diagrams for magnetic field strengths
close to the anticrossings. Figure~\ref{fig:CB_wc=2.1} shows the
current (upper part) and the differential conductance (lower part) for
$B=7.24$~T.  For small transport voltages, the three-electron GS is
only weakly polarized and the GS-GS transition between two and three
electrons is allowed. Therefore, Fig.~\ref{fig:CB_wc=2.1} shows
current already for small transport voltages in contrast to
Fig.~\ref{fig:CB_wc=2.245}. But increasing the source-drain voltage
also increases the asymmetry between the dots and therefore the
polarization of the three-electron GS, which reduces the spectral
weight for the tunneling-in process.  Hence the current may decrease
with increasing transport voltage giving rise to negative differential
conductances.

In conclusion: We observed and explained correlation induced parity crossings in
a symmetric DQD in a magnetic field sweep. These crossings convert to a
magnetic-field-dependent charge-localization for an odd numbers of
electrons inside the DQD if the symmetry between the dots is broken.
This localization strongly affects the transport characteristics and
leads to an isospin blockade as well as to negative differential
conductances.
We thank Michael Tews for many illuminating discussions and
acknowledge financial support by the Deutsche Forschungsgemeinschaft
via SFB 508.


\begin{thebibliography}{9}
\bibitem{material} Throughout this
  paper we have used the GaAs parameters for the effective mass $m^*$
  of the electrons and the Land\'{e} factor $g^*$, e.g. $m^*=0.067m_e$
  and $g^*=-0.44$.
\end{thebibliography}
\end{document}